**A dimensão espacial das fases da Lua: contribuições para uma proposta de ensino**

*Alejandro Gangui*
*Esteban Dicovskiy*
*Maria C. Iglesias*

1. **Introdução**

Neste capítulo, apresentamos algumas reflexões acerca da aprendizagem e sua implicação no ensino das noções sobre as fases da Lua, que surgem como consequência de investigações implementadas por nosso grupo de trabalho. Reconhecemos como valiosa a necessidade de prestar atenção à descrição do fenômenos a partir de um ponto de vista sobre a superfície terrestre. Nesse contexto, refletimos a respeito da dificuldade associada à tridimensionalidade do Sistema Sol-Terra-Lua, uma vez que consideramos a linguagem iconográfica, e discutimos acerca do uso de esquemas e representações deste fenômeno, assim como a respeito das imagens e fotografias da Lua. Por fim, apresentamos algumas considerações para o ensino desse tema.

Imagine que temos que preparar uma aula sobre as fases da Lua. Quais são os aspectos que focaremos? Certamente, buscaremos fazer com que os alunos possam estabelecer qual é a posição relativa dos três corpos (Sol-Terra-Lua), para que ocorram as diferentes fases lunares: a Lua nova, crescente, cheia e minguante. Para isso, podemos usar algum esquema, muitas vezes, presente em diversos livros didáticos, onde são apresentados o Sol e a Terra fixos, e *várias* Luas ao redor de nosso planeta (fig. 1).

Figura 1: Imagem que pretende representar o fenômeno das fases da Lua, visto a partir de uma posição localizada no espaço exterior. A presença do Sol, no desenho, confunde o leitor e faz com que a iluminação das Luas não seja correta. A Terra também não possui a iluminação esperada, considerando o arranjo dos astros envolvidos. Esta imagem apresenta, ao mesmo tempo, a observação do fenômeno a partir da Terra (sem explicitar em qual hemisfério está o observador) e de um ponto de vista externo a ela. Por seu caráter bidimensional, no plano, o esquema não consegue explicar este fenômeno espacial.

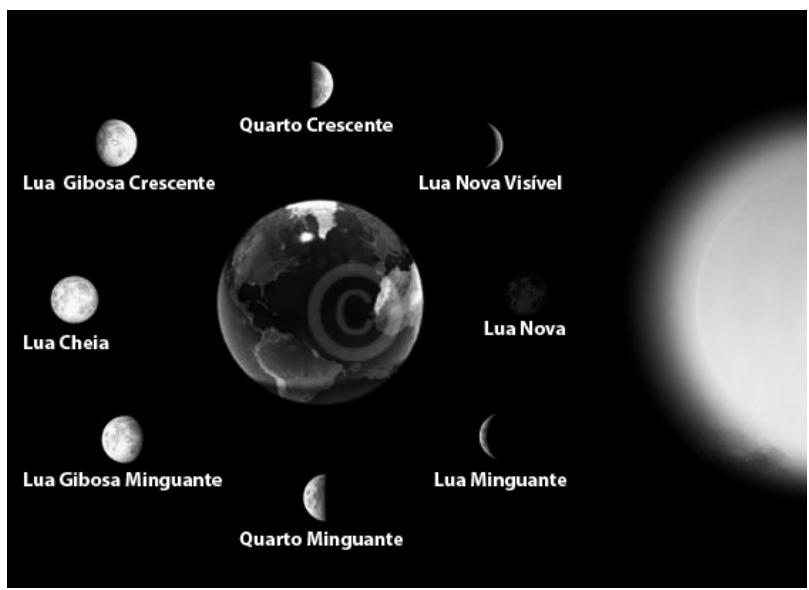

Ou seja, para desenvolver estes conteúdos, possivelmente, o façamos de uma perspectiva centrada em um ponto de vista *externo* a Terra, típico de um observador situado no espaço exterior (e muito longe do plano da órbita da Lua em torno da Terra). Além disso, buscaremos com que os alunos reconheçam que a Lua gasta um pouco menos que um mês para transladar ao redor da Terra, e que seus períodos de translação e rotação sobre seu eixo são, aproximadamente, iguais.

Ainda que tais ideias sobre a Lua sejam importantes, nós reconhecemos que elas *não são suficientes* para resolver situações que envolvem o sistema Sol-Terra-Lua (STL). Acontece que muitos estudantes podem, sem maiores inconvenientes, reproduzir típicos esquemas dos livros didáticos, mas não conseguem, por exemplo, indicar com precisão o quanto a Lua se afasta do Sol, na abóbada celeste, a cada dia, nem como será a evolução da uma determinada fase lunar após alguns dias, no decorrer da semana. Sabemos, também, que, com o passar das horas, nosso satélite natural muda de aspecto, a orientação de sua parte iluminada, fazendo com que o simples desenho de uma determinada fase deva ser submetido a reflexões para compreender plenamente o observado. A grande maioria dos alunos não se questiona sobre tais temas, nem compreende as razões para a variabilidade das fases lunares.

Para começar a pensar sobre essas questões, lembremos que o *ponto de referência* a partir do qual fazemos as observações está localizado na superfície da Terra. Posto que nossos sentidos nos indicam que todos os objetos no céu parecem se mover em torno de nós, cria-se o espaço ideal para a construção de algumas ideias intuitivas. Existem, também, outros obstáculos de aprendizagem que se mostram resistentes, interferindo na construção de conhecimento mais científico (GANGUI, IGLESIAS e QUINTEROS, 2010). Podemos citar, por exemplo, o problema da linguagem cotidiana, que reforça algumas das ideias que os alunos trazem, ou o uso de maquetes sem escala, ou quando, sem nos darmos conta, impomos aos estudantes a explicação que a ciência oferece, atualmente, sobre as fases da Lua, sem oferecer um espaço de reflexão sobre o assunto. No entanto concordamos em que não é simplesmente reconhecer qual é a explicação científica atual, mas compreender como é que as observações e os dados se compatibilizam e se encaixam perfeitamente nela. Certamente, os alunos terão várias explicações possíveis para as fases da Lua, e algumas destas respostas poderiam, inclusive, explicar satisfatoriamente os fenômenos investigados. Então, o que se pode destacar aqui é a importância de se refletir sobre se as ideias propostas permitem justificar as alterações e as permanências observadas (MARTÍNEZ SEBASTIÀ, 2004). Assim mesmo, devemos ter em mente que todos esses elementos são essenciais para a construção de um determinado modelo de explicação.

Tal como acontece com a todalidade dos fenômenos astronômicos observáveis a olho nu, as fases da Lua têm características próprias, que dependem do lugar onde a observação é realizada. No entanto poucas são as vezes em que partimos de nossas *vivências astronômicas* e fazemos referência a elas. Além disso, muitas vezes, usamos imagens e esquemas nos quais se representam, de maneira bidimensional, o espaço tridimensional. Todos estes aspectos podem se tornar dificuldades de aprendizagem e levar a uma visão distorcida de como se produzem as fases lunares. A isto, podemos acrescentar o fato de, novamente, sem nos darmos conta ou pensarmos a respeito, usarmos livros, imagens ou vídeos que descrevem situações do céu que não são comuns em nossas latitudes, mas que correspondem a observações realizadas em áreas geográficas muito distantes da nossa localização sobre a superfície da Terra.

## 2. As explicações sobre as fases da Lua

Em trabalhos recentes, nosso grupo de pesquisa pôde contribuir para o diagnóstico da situação de futuros professores, que estudam nos Institutos Superiores de Formação Docente da Cidade de Buenos Aires, no que diz respeito a temas de Astronomia. Em particular, nossa atual linha de trabalho lida com as representações que eles têm sobre o fenômeno das fases da Lua e questiona sobre seus diferentes modelos explicativos e acerca das dificuldades subjacentes à compreensão significativa do fenômeno, além de suas correspondência com o modelo científico aceito na atualidade (IGLESIAS et. al, 2012;. DICOVSKIY et. al, 2012).

A partir da análise das respostas dadas pelos alunos avárias perguntas, revelamos as dificuldades que surgem quando lhes pedimos para resolver uma situação que se refere ao fenômeno das fases, observado partir da Terra, mas que requer uma explicação externa, como é o caso da Lua nova. Poucos são os estudantes que encaminham algum tipo de explicação, seja cientificamente aceita ou alternativa. Pelo contrário, um significativo número de estudantes não trazem explicação para o fenômeno em discussão, mas, sim, oferecem respostas do tipo descritivas e baseadas em suas vivências. Curiosamente, a maioria de estudantes que não apresenta uma explicação, traz, em troca, uma descrição sobre o que é a Lua nova ou o que se vê no céu, quando nosso satélite natural está nessa fase. Por exemplo, vários alunos afirmam, explicitamente, que a Lua, nesse dia, não tem iluminada a "cara" que vemos a partir da Terra ou que não é percebida no céu, mas não especificam a quê se deve este fenômeno. Por outro lado, geralmente, há uma clara referência à noite, seja em seus escritos ou quando os complementa com um desenho, no qual são incluídas estrelas bem visíveis. Esta associação da Lua com a noite também foi descrita por Samarapungavan, Vosniadou e Brewer (1996), em trabalho no qual se destaca que nenhuma criança da Índia, que foi entrevistada, parece aceitar que a Lua pode, realmente, ser vista durante o dia.

No geral, podemos dizer que ante a apresentação aos alunos investigados de uma Lua em particular, como a Lua nova, é difícil a eles vincular essa observação com o fenômeno das fases lunares ou fazer uso de uma explicação das fases para dar resposta a um caso concreto e percebido a partir da Terra. Sobre estas questões, Martínez Sebastià (2004) expõe a dificuldade revelada por suas pesquisas sobre o sistema Sol-Terra-Lua, nas quais comprovou que os alunos evidenciam dificuldades em separar o conhecimento fenomenológico do conhecimento teórico do próprio modelo.

Finalmente, descobrimos, em nossas investigações, que uma porcentagem elevada de estudantes pesquisados fazem desenhos que ilustram o escrito ou, no máximo, incluem elementos que chamamos de "infantis" (Fig. 2).

Figura 2: Diante de uma situação relacionada com a incapacidade de ver a Lua nova a partir da Terra, o investigado inclui elementos desnecessários para resolver a situação em questão. Desenho de um aluno no contexto de uma pesquisa educacional.

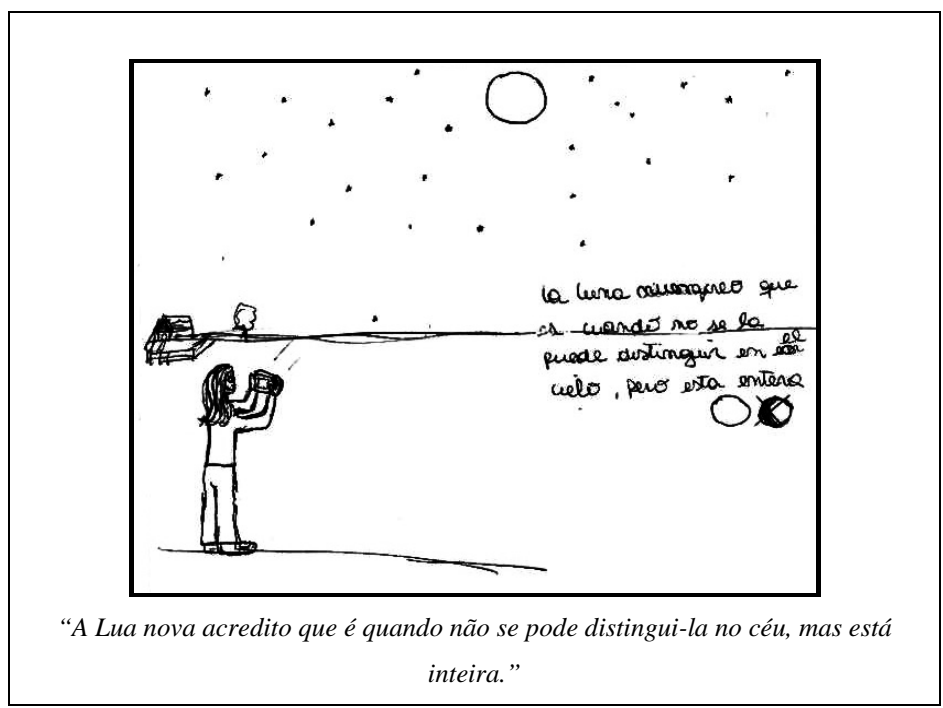

*"A Lua nova acredito que é quando não se pode distingui-la no céu, mas está inteira."*

É interessante destacar que a possibilidade de trabalhar com materiais concretos, tais como esferas de isopor e fontes de luz, proporcionou mudanças importantes. Forçou os alunos a colocar de lado suas primeiras respostas vivenciais ou caracterizadas por elementos infantis, a sair da posição de observador terrestre e avançar na direção de um tipo diferente de resposta, ou seja, que relacionava, espacialmente, os três corpos envolvidos (STL) e com algum grau de coerência, embora não inteiramente correta.

Em geral, quando pedimos aos alunos para usar os elementos disponíveis para pensar a quê se deve o fato de não podermos ver a Lua quando ela se encontra na fase nova, os investigados, geralmente, respondem que a Terra se coloca entre a Lua e o Sol, produzindo sombra. Se pedirmos a eles que façam um desenho para representar a situação descrita, os estudantes, muitas vezes, colocam os três astros alinhados (Fig. 3).

Figura 3: Típico desenho produzido por alunos, quando solicitados a explicar como a Lua nova ocorre. A literatura específica chama esta situação de "teoria do eclipse", uma vez que os astros envolvidos estão alinhados. Observemos como, a partir desta explicação, a Lua fica oculta no "cone de sombra" da Terra.

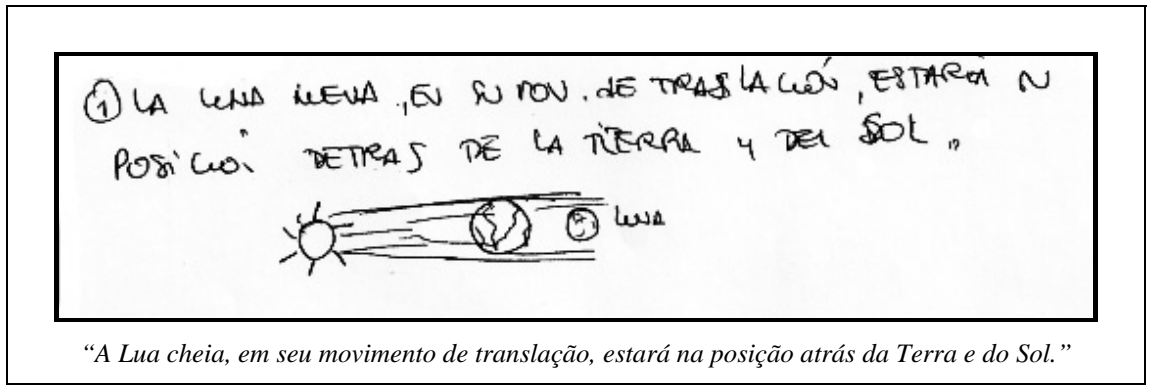

*"A Lua cheia, em seu movimento de translação, estará na posição atrás da Terra e do Sol."*

A ideia de sombra produzida pela Terra não é completamente ilógica, se nos lembrarmos de que todo corpo opaco gera uma sombra com sua mesma forma, quando é iluminado por uma fonte de luz; uma explicação que funciona muito bem para muitos fenômenos cotidianos. A noção que se atribui às fases lunares a sombra da Terra é predominante em todas as idades, existindo uma grande confusão entre a explicação das fases lunares a dos eclipses lunares (BAXTER, 1989; SCHOON, 1992; IACHEL, LANGHI e SCALVI, 2008). Em alguns desses estudos, tornou-se claro que essa noção parece ser mais comum em estudantes mais velhos, sendo elevada para alunos que continuam seus cursos de ensino secundário (SCHOON, 1992).

Ao contrário do que aconteceu no caso em que as propostas eram claramente situações vivenciais, quando, em nossa pesquisa, pedimos aos estudantes para usar os elementos disponíveis para explicar como se produz outra fase lunar, por exemplo, a Lua cheia, em geral, eles se encontram ante a uma situação de conflito que os leva a pensar em novas explicações. Isso evidencia que os pesquisados entendem que suas primeiras respostas para os mesmos problemas, agora, parecem incertas:

• Um grupo proporcionalmente grande de estudantes propõe posições "diametralmente" opostas. De posicionar a Lua no "cone se sombra" da Terra, o que reforça as respostas das questões iniciais, e passa logo a colocá-la do outro lado da Terra, em frente ao Sol, de modo que, assim, fique iluminada (Fig. 4). Verifica-se que eles não se deparam com qualquer conflito pelo fato de a Lua resultante *não* ser cheia, mas, sim, nova, e tampouco o fato de que, para ver a Lua nessa posição, terá que posicionar-se ao lado do Sol. Além disso, não percebem que, em posições como estas, novamente, são produzidos eclipses quinzenais.

Figura 4: Exemplo de resposta que os alunos apresentam no contexto de uma pesquisa educacional, quando é pedido a eles que expliquem como ocorre a Lua cheia. Para resolver o conflito de onde localizar a Lua, para que fique totalmente iluminada (e evitar que caia no cone de sombra da Terra), os investigados a colocam entre a Terra e o Sol, alinhando os três astros.

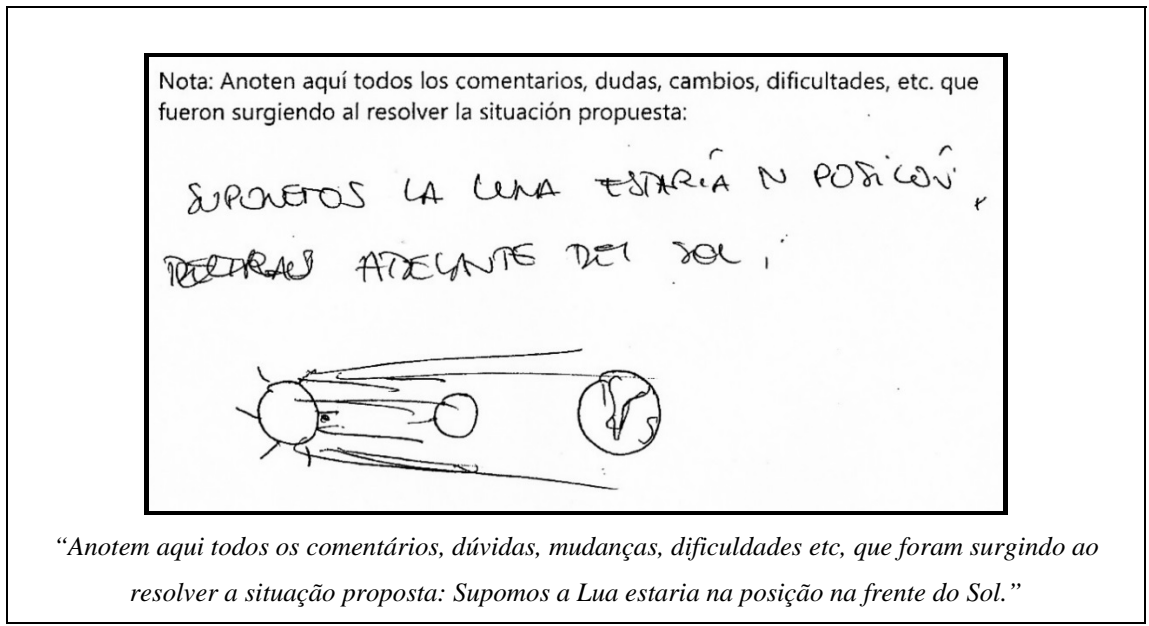

*"Anotem aqui todos os comentários, dúvidas, mudanças, dificuldades etc, que foram surgindo ao resolver a situação proposta: Supomos a Lua estaria na posição na frente do Sol."*

• Outro grupo de alunos, que, inicialmente, também indicavam que a Lua nova ocorria por cair dentro da sombra produzida pela Terra, ante o problema da fase cheia, coloca a Lua deslocada desta área escura, do mesmo lado de nosso planeta onde estava antes, ainda que "deslocada um pouco". Desta forma, esses alunos superaram a situação das respostas anteriores e atentam para o fato de que assim se poderá, efetivamente, ver a Lua a partir da Terra. Estes alunos respeitam seu movimento no plano da órbita da Terra, assumindo a bidimensionalidade do movimento destes corpos (Fig. 5).

Figura 5: Exemplo de resposta que os alunos apresentam no contexto de uma pesquisa educacional, quando são solicitados a explicar como ocorre a Lua cheia. Para resolver o conflito sobre onde colocar a Lua para que esteja totalmente iluminada, decidem tirá-la do "cone de sombra".

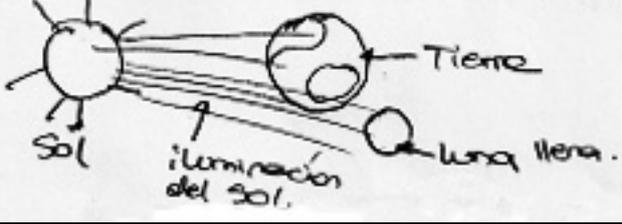

*"Anotem aqui todos os comentários, dúvidas, mudanças, dificuldades, etc, que foram surgindo ao resolver a situação proposta: Tivemos dúvida, primeiro, na posição que deveria ter a Lua para que fosse Lua cheia. Logo, sobre como se movia. Cremos que a Lua gira de forma oblíqua ao redor da Terra, e é a Lua cheia quando da Terra se vê toda a face iluminada."*

Notavelmente, são poucos os alunos que apresentam respostas em que reconhecem que lhes gera conflito o fato de que o fenômeno ocorra apenas no plano da órbita da Terra e, diretamente, apresentam a necessidade de se pensar um movimento do sistema STL em três dimensões, o que os aproxima de respostas mais científicas. Nesse pequeno grupo, encontramos tentativas de "romper" com a linearidade, movendo a Lua e destacando a necessidade da tridimensionalidade ou a necessidade de perspectiva (Fig. 6).

Figura 6: Exemplo de resposta apresentada pelos alunos no contexto de uma pesquisa educacional, quando solicitados a explicar como ocorre a Lua cheia. Para resolver o conflito acerca de onde colocar a Lua para que fique totalmente iluminada, decidem posicioná-la do mesmo lado que o Sol, tentando "romper" com a linearidade dos astros. Desenham, ainda, a localização que deve ter o observador, neste caso, representado por uma personagem chamada "Juliana".

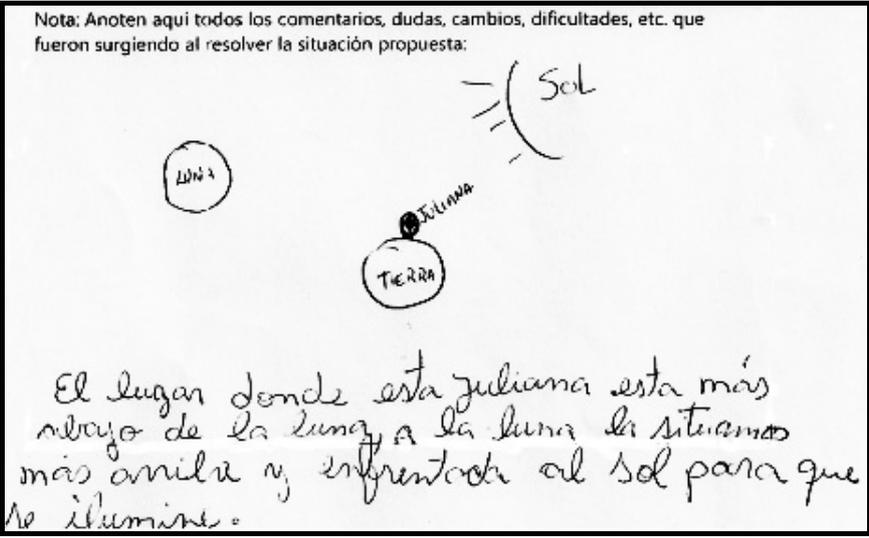

*"Anotem aqui todos os comentários, dúvidas, mudanças, dificuldades, etc, que foram surgindo ao resolver a situação proposta: o lugar onde está Juliana está mais abaixo da Lua. Situamos a Lua mais acima e em frente ao Sol para que se ilumine."*

Como podemos reconhecer, com base nessa breve revisão de respostas que surgem no contexto de uma pesquisa educacional, os maiores problemas são causados por um forte apego à visão *topocêntrica* dos fenômenos celestes, isto é, devido à perspectiva de observar o céu de um determinado lugar da superfície da Terra.

Voltemos, agora, a pensar no ensino e aprendizagem das fases da Lua. Será possível obter todos os conceitos discutidos, só pedindo ao aluno que se lembre da posição da Lua, da Terra e do Sol para cada fase lunar?

### 3. A visão espacial e os conhecimentos prévios

Resolver situações-problema que envolvem as fases da Lua requer um alto nível de abstração, uma vez se que supõe poder estabelecer diferentes perspectivas para sua observação: (1) aquilo que vejo a partir da minha posição topocêntrica, (2) aquilo que veriam os outros observadores terrestres, e (3) aquilo que vejo quando me afasto, imaginariamente, da Terra. Embora este nível de abstração também esteja presente quando resolvemos situações que envolvem o sistema Sol-Terra, no caso das fases da Lua, são três corpos de uma só vez, o que supõe um grau de abstração maior. Além disso, como afirmamos, requer uma certa prática pensar e trabalhar em três dimensões. Neste sentido, torna-se essencial levar em consideração as dimensões dos astros envolvidos, tanto seus tamanhos relativos quanto as distâncias e seus movimentos.

Muitas vezes, acontece que as imagens ou modelos que usamos nas aulas de ciências para tornar mais compreensível o tema, não respeitam tais aspectos e, em outras ocasiões, eles não são analisados com nossos alunos. Algo semelhante pode acontecer, quando usamos esferas de isopor, sem respeitar as dimensões espaciais (por exemplo, as magnitudes características dos astros), uma vez que os alunos poderiam aplicar, literalmente, alguns aspectos de simulação ao fenômeno, como, por exemplo, a distância entre os astros ou seus tamanhos (LANCIANO, 1989). Devemos nos lembrar de que por meio do uso de representações concretas, buscamos que os alunos possam pensar, dizer e fazer a respeito do mundo dos fenômenos sobre os quais estão trabalhando (CAMINO, 2004). Neste sentido, os modelos concretos devem ser instrumentos que acompanhem os alunos no processo de ressignificação de suas vivências astronômicas cotidianas, sem se esquecer de que elas estão acompanhadas de ideias prévias.

### 4. O movimento aparente e as mudanças na iluminação da Lua

Destacamos o valor didático que possui a observação das fases a partir do ponto de vista um observador qualquer, tomando como referência sua posição topocêntrica. Apresentamos, agora, de forma resumida, o acompanhamento da Lua que pode ser feito no céu, em relação ao Sol.

Lembremos que a Lua, assim como o Sol, possui um movimento aparente pela abóbada celeste e acima do nosso horizonte. Para este estudo, podemos descrever três importantes momentos de sua trajetória no céu: (1) seu despontar, (2) seu ponto mais alto ou culminação e (3) seu poente ou ocaso (Fig. 7).

Figura 7: Esquema que representa o movimento aparente da Lua. A parte cinza da Lua corresponde à parte iluminada. Os números 1, 2 e 3 representam três momentos aproximados e distintos de seu caminho aparente sobre o horizonte. O esquema corresponde a um típico observador do Hemisfério Sul. Para observadores do Hemisfério Norte, o esquema deveria voltar-se para o ponto cardeal Sul; a posição 1 corresponderia, aproximadamente, ao ocaso e a 3 ao seu despontar pela manhã.

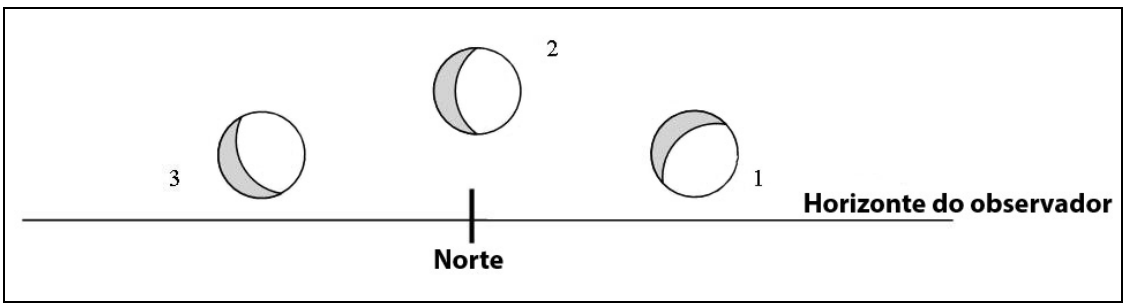

Vamos supor que queremos saber, para a nossa latitude, a que hora a Lua sai no quarto crescente, em um determinado mês. Consultamos, então, uma tabela de efemérides astronômicas, a qual nos informa que o horário de saída é às 12h (hora solar). Nesse instante, onde está o Sol? Para simplificar a situação, assumimos que estamos em um dia perto de um dos equinócios, quando a duração do caminho aparente do Sol é de cerca de doze horas. Se supusermos que o amanhecer do dia foi às 6h (hora solar), isso quer dizer que, quando a Lua crescente estiver começando a aparecer no horizonte, o Sol vai estar na metade de seu caminho aparente (Fig. 8). Em outras palavras, estamos justamente no meio-dia solar.

Figura 8: Esquema que representa a saída da Lua no quarto crescente pelo horizonte oriental e sua relação com a posição do Sol, no céu, para um típico observador do Hemisfério Sul.

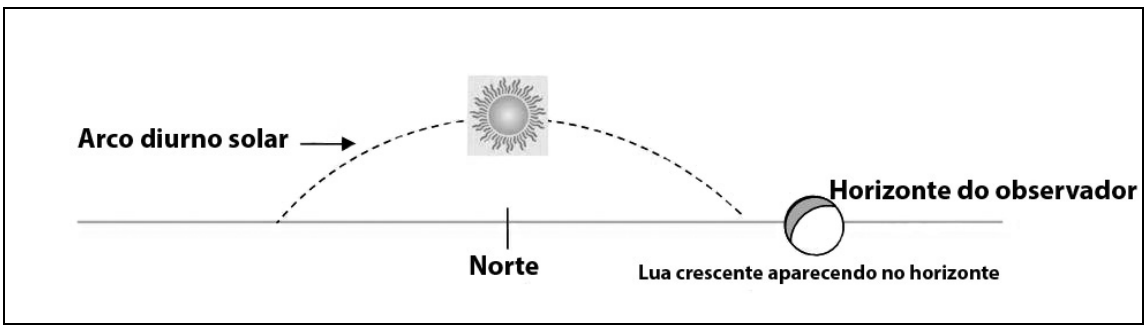

Na configuração mostrada na figura anterior, com o Sol alto no céu, ao meio-dia, vemos que a Lua, quando desponta no horizonte, é iluminada pelo Sol "por cima". Com

o passar das horas, tanto a Lua quanto o Sol seguem em sua trajetória aparente. Quando nosso satélite natural começa a se aproximar do seu ponto de culminação, o Sol já está próximo ao ocaso (Fig. 9), no quadrante ocidental, de tal modo que a Lua terá a iluminação típica das imagens que vemos em muitos livros didáticos.

Figura 9: Esquema que representa o ponto de culminação da Lua em seu caminho aparente pelo céu diurno e sua relação com a posição do Sol, para um típico observador do Hemisfério Sul. Nesta situação astronômica particular, recupera-se a imagem da Lua que, geralmente, encontramos em muitos livros didáticos, isto é, a Lua iluminada formando um "C" (muitas vezes, relacionado com o C de "crescente").

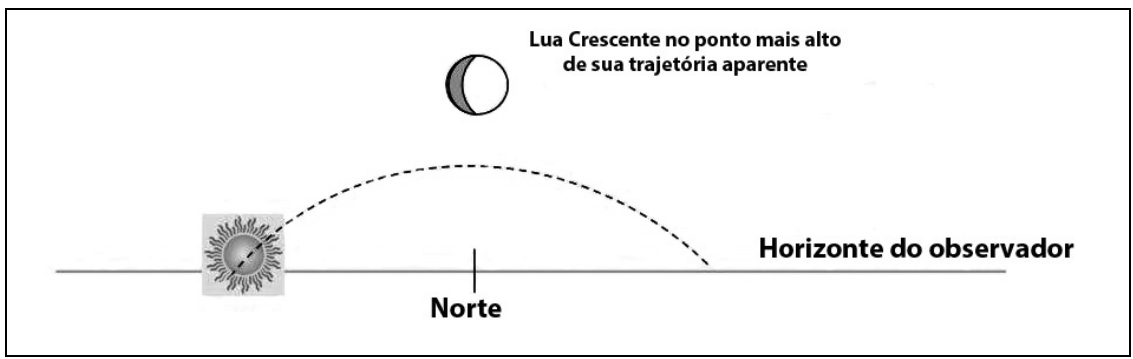

Como é a iluminação da Lua, quando ela está se aproximando do ocaso? Naquele momento, o Sol já não é visível para nós, uma vez que vai estar abaixo do horizonte. Nesse caso, ele iluminará a Lua "por baixo" (Fig. 10).

Figura 10: Esquema que representa os minutos que antecedem o ocaso da Lua, no horizonte ocidental, em sua fase de quarto crescente, para um típico observador do Hemisfério Sul.

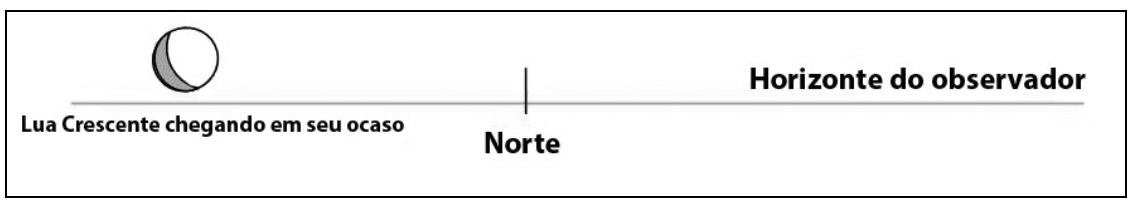

Embora os esquemas apresentados aqui estejam muito simplificados, eles nos permitem visualizar os movimentos aparentes e relacionar ambos os astros de uma só vez, elementos importantes para uma compreensão adequada das fases da Lua. Poderíamos também fazer outros esquemas e análises como os apresentados aqui, mas para outros momentos do ciclo lunar. Para isso, poderemos usar os dados de nascer e pôr da Lua, fornecidos pelas tabelas de efemérides, ou mesmo alguns simuladores astronômicos, livremente gratuitos na internet.

## 5. A Lua em livros didáticos e na Internet

Nas seções anteriores, consideramos a importância de realizar o acompanhamento sistemático da Lua em relação ao Sol. Seja por falta de espaço, tempo ou disponibilidade, a observação do céu, nem sempre, é fácil, especialmente, em grandes cidades, onde o horizonte não é visível. Para corrigir isso, podemos recorrer ao uso de uma coleção de imagens que podem ser encontradas em sites da internet, ou aqueles que aparecem nos livros didáticos (Fig.11). Mas quais são os aspectos que devemos levar em consideração na hora de selecioná-las? O que estas imagens nos dizem sobre a Lua e suas fases?

Figura 11: Coleção de imagens de diferentes fases da Lua extraídas de um livro didático.

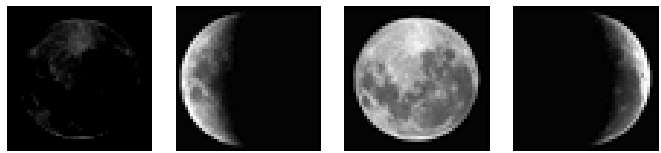

Geralmente, quando queremos fazer uma coleção de imagens da Lua, damos especial atenção à iluminação da face visível de nosso satélite e selecionamos uma variedade delas, de forma a termos representadas todas as fases possíveis.

O que um aluno poderia dizer a respeito do momento do dia em que foram obtidas? Sabemos que uma das ideias previas mais comuns entre os alunos é aquela que associa a noite com a Lua. O uso excessivo de Luas com *fundo escuro* poderia reforçar tal ideia e interferir na construção de noções sobre o tema. Como vimos anteriormente, o Sol e a Lua podem compartilhar o céu *diurno*, pelo menos durante algum tempo de suas trajetórias aparentes. Observando os horários de saída da Lua durante vários dias consecutivos, aprendemos que ela demora mais que o Sol para aparecer a cada manhã, e, cada dia, demora mais. O Sol, no entanto, modifica em um minuto, aproximadamente, sua aparição matinal, dia após dia. Essas diferenças entre ambos os astros nos permitem supor que, durante o ciclo lunar, haverá momentos em que a Lua "acompanha" o Sol em sua jornada pelo céu. Por exemplo, como mostrado na tabela de efemérides (Tabela 1), no dia 8 de setembro de 2012, a Lua cruzou o horizonte às 01h26 da madrugada[1]. O Sol, no entanto, ainda estava abaixo do horizonte. Se, nesse dia, esperarmos cerca de 6 horas, veremos que o Sol estará prestes a despontar no horizonte, enquanto a Lua vai ser encontrar perto do ponto mais alto de sua trajetória aparente.

Tabela 1: Dados com a hora civil do despontar, do trânsito e do poente da Lua (na fase minguante) e sua relação com o Sol, para o dia 8 de setembro de 2012, na cidade de La Plata (coordenadas: 34° 54' 30" S, 57° 55' 54" W). Nota-se que, por mais de quatro horas e meia (das 7 às 11h40, aproximadamente), ambos os astros compartilham o céu. Este período de tempo vai aumentando, à medida que nos aproximamos da fase de Lua nova.

| Astro | Horário Nascente (horas e minutos) | Horário Trânsito[2] (horas e minutos) | Horário Poente (horas e minutos) |
|---|---|---|---|
| **Lua** | 01:26 | 06:34 | 11:40 |
| **Sol** | 07:01 | 12:49 | 18:38 |

O que podemos concluir? Pelo menos para uma parte do seu curso aparente, o Sol e a Lua, neste caso, em fase descrescente ou minguante, compartilharam do céu. Como podemos ver que o Sol está presente no céu, então, não pode ser noite e, portanto, a imagem que representa a Lua não pode ter fundo escuro.

Nas seções anteriores, comentamos a respeito das dificuldades encontradas nos alunos no momento de dar explicações para as fases da Lua, as quais estão ligadas à compreensão de que o fenômeno tem caráter topocêntrico e, portanto, sabemos que a posição do observador é um elemento central para sua correta explicação. Outro aspecto que interfere com a explicação tem a ver com o conceito de tridimensionalidade e a periodicidade do fenômeno, dado que as explicações mais comuns ficam, erroneamente, circunscritas a configurações lineares ou bidimensionais do Sol, da Terra e da Lua.

Na verdade, é muito raro que se dê o passo necessário para alcançar explicações tridimensionais, ou até mesmo que se chegue a notar sua necessidade. Nestas circunstâncias, é necessário considerar de que forma as imagens, como recurso amplamente utilizado em sala de aula, podem se tornar um meio facilitador da comunicação científica no ensino, ou tomar a direção oposta. Os esquemas que, normalmente, são apresentados em livros didáticos para explicar as fases da Lua o fazem em *duas dimensões*, o que *obstaculariza* a compreensão das diferenças importantes que existem entre as fases e os eclipses. Da mesma forma, as frases, muitas vezes, usadas para descrever a posição de objetos celestes, no caso de descrever a Lua nova ou a cheia, como, por exemplo, "se coloca entre" ou "está entre", poderiam reforçar as ideias alternativas que os alunos apresentam, uma vez que, implicitamente, fazem alusão à linearidade na distribuição espacial dos astros (KARASEUR, 2012).

---

[1] Os instantes são dados na horário oficial da Argentina e correspondem à cidade de La Plata. Os dados são provenientes da Faculdade de Ciências Astronômicas e Geofísicas (U.N. de La Plata).
[2] O momento de máxima altura sobre o horizonte, ou seja, o instante em que o astro cruza o meridiano do lugar, é chamado de trânsito ou culminação.

### 6. As fases da Lua: considerações para o ensino

Se tivermos em mente as dificuldades que surgem, quando queremos que nossos alunos aprendam as noções relacionadas ao tema das fases da Lua, é importante considerar propostas de ensino que levem em conta a forma progressiva de organização dos conteúdos. Acreditamos que as fases da Lua seja um tema que, em princípio, pode ser estudado a partir de uma visão situada na superfície terrestre e sem necessidade de deixar a Terra. Assim, será conveniente deixar a explicação das fases descritas a partir de um ponto de referência externo a Terra, para quando nossos alunos já estejam familiarizados com o fenômeno, de maneira a poder relacionar de forma progressiva ambas as visões.

É importante lembrar que, no momento de planejar uma sequência de ensino, será necessário certificar-se de que os alunos conhecem algumas das principais características da Lua, entre elas, que é um satélite natural da Terra e que suas fases correspondem a uma parte iluminada pelo Sol de sua face visível.

### 6.1 Conteúdos escolares: potenciais noções como objetivos

Na maioria das propostas curriculares em que se apresenta o estudo das fases da Lua, os conteúdos a serem ensinados têm como objetivo poder aproximar os alunos das explicações científicas, que se referem a observações do fenômeno em uma posição de fora da Terra e de uma noção de espaço tridimensional. Se levarmos em conta as limitações decorrentes das considerações acima expostas, é válido repensar o que é que esperamos que os alunos aprendam, mas com foco no reconhecimento desse fenômeno a partir de uma posição topocêntrica. Será importante determinar, então, que noções poderiam ser propostas para serem ensinadas em uma potencial sequência de ensino sobre as fases da Lua, que, geralmente, não são especificadas nas propostas curriculares, que só ficam limitadas à observação das fases como imagens fixas da Lua para cada momento do mês lunar.

Como uma primeira aproximação, podemos levantar os seguintes pontos:

• Reconhecer que a Lua é observada em diferentes momentos do dia, tanto nas horas diurnas quanto durante à noite, e que *não* é mais provável observá-la à noite do que de dia;
• Identificar como são as mudanças de fases da Lua, dia a dia, semana a semana;
• Identificar e comparar os horários de saída, de culminação e ocaso de ambos os astros;
• Identificar e reconhecer os movimentos aparentes do Sol e da Lua, sua posição no céu e ao longo do tempo;
• Identificar como vai mudando, dia a dia, a distância entre ambos os astros, no céu, de acordo com os movimentos e horários de saída;
• Reconhecer como muda a iluminação da Lua a cada dia, e sua relação com a posição do Sol no céu;
• Identificar e comparar os "arcos" aparentes que ambos os astros descrevem no céu.

**6.2 Introdução ao tema de estudo: a problematização**

Ressaltamos a importância de implementar algumas perguntas "disparadoras" que convidem os alunos a lançar suas ideias sobre o tema, a fim de conhecer seus conhecimentos prévios e, assim, revelar suas explicações. Isto oferece a possibilidade de encontrar respostas contraditórias sem aparente conflito para os alunos. Por outra parte, a definição de problemas relevantes e consistentes com as noções que se quer ensinar implica o percurso didático que deve ser levado adiante, e à medida que os alunos participam deste problema, estarão mais envolvidos no estudo dessas noções.

A aparição da Lua no céu diurno, em *vários* dias do mês lunar (na verdade, em *quase todos* os dias do mês, exceto, talvez, durante o dia correspondente à Lua cheia), e a comparação desta situação com as imagens das "28 fases" da Lua (com exceção do dia que corresponde, precisamente, à Lua nova), todos com fundo de céu noturno, podem se constituir em valiosos recursos para a apresentação de um problema sobre este fenômeno.

Também é possível propor aos alunos que determinem o lugar onde o Sol deve ser encontrado para justificar as observações da Lua em diferentes fases, perguntando como é possível que ambos os astros possam se encontrar em lugares tão diferentes do céu, se o lógico é pensar que seus movimentos quase não variam dia após dia.

Problemas desse tipo, em que se problematizam observações ou situações que são familiares e cotidianas, obriga os alunos a empregar seus conhecimentos e confrontá-los com novos olhares sobre esse fenômeno, estimulando neles a análise e a busca por novas explicações que lhes permitam esclarecer e avançar sobre as questões propostas.

**6.3 Instâncias de observação e registro**

O estudo dos fenômenos das fases da Lua e as noções ligadas a ele, principalmente, centrados em uma perspectiva topocêntrica, determina que os procedimentos centrais em uma seqüência de atividades sejam o da observação e registro. Isto é, na medida do possível, coloca a necessidade de observações a olho nu da Lua e sua relação com o Sol, para a posição do observador.

Durante as primeiras saídas de observação, pode ser desejável rever ou introduzir alguns conceitos que ajudarão na interpretação desses fenômenos. Por exemplo, é possível familiarizar-se com a forma da abóbada celeste que parece ter o céu dos observadores, bem como especificar a posição dos astros no céu, o que é útil para introduzir as coordenadas locais de altura angular e azimute[3].

Somando-se a essas primeiras considerações, citamos potenciais observações e registros sobre aqueles elementos que permitem a interpretação e a análise dos fatos que ocorrem em torno das fases da Lua:

• *A importância de identificar e registrar como se encontra iluminada a face da Lua em diferentes momentos do ciclo lunar. Levar adiante estas simples observações iniciais permitirá ao aluno familiarizar-se com a tarefa a realizar.*

Seguir esse fenômeno ao longo de um dia em que se identifique seu movimento aparente será supor que ela se move dentro de um espaço em formato de abóbada. Além disso, poderá se perguntar qual é a localização do Sol, em cada caso, que determina qual parte da Lua está iluminada.

---

[3] A altura angular é a distância angular medida a partir do horizonte até o astro. Ela é medida em graus, perpendicular ao horizonte do observador. É positiva, se o astro se encontra no hemisfério visível, e negativa, se está abaixo do horizonte. O azimute é a distância angular medida no horizonte a partir do Norte, em direção ao Leste. Possui medidas entre 0 e 360 graus.

• *O valor de esquematizar o movimento aparente da Lua e do Sol. Ao comparar o caminho de ambos os astros pela abóbada celeste, é possível explorar alternativas e ver que, em geral, suas trajetórias aparentes não coincidem.*

Essa análise permitirá discutir com os alunos se os arcos das trajetórias da Lua e do Sol são ou não os mesmos. Esclarecer esta situação favoreceria questionar como veríamos os astros, especialmente, a Lua, se ambos os arcos coincidissem.

• *A importância de reconhecer mudanças na iluminação do céu ao observar a Lua em suas diferentes fases, ao longo do tempo (ao longo dos dias e também durante as horas do mesmo dia). Contrastar a cor do céu por "detrás da Lua" em diferentes dias e horários permite reconhecer em que momento ocorre uma dada fase de acordo com a posição do Sol e daqueles que o observam.*

Diferenciar entre somente perceber a Lua no céu e examinar *sua presença* no céu, de acordo com estas observações, ganha força ao comparar com as imagens que, muitas vezes, aparecem de forma imprecisa em livros, onde se associa a Lua com o céu noturno e como algo pouco frequente de ser ver durante o dia. Além disso, a observação e o registro de tais situações deveriam ajudar a analisar que, ainda que não vejamos a Lua em um céu azul, ela pode estar, sim, no céu, ainda que muito tênue e difícil de localizar.

As interpretações e análises dessas observações e seus registros serão enriquecidas se houver fotos reais da Lua em diferentes momentos para contrastar com suas respostas. Do mesmo modo será de grande valor contar com *softwares* de simulação[4], que não só permitem uma comparação com os registros realizados, mas que facilitam também o reconhecimento de que estas observações ocorrem e mudam ao longo de um período de tempo.

Os recursos que acabamos de mencionar se transformam em meios fundamentais de observação, no caso de não se poder contar com a possibilidade de realização de observações diretas do céu ou que, por excesso de construção na cidade, fique restrito a alturas angulares maiores, só próximas ao zênite.

---

[4] Entre os simuladores do céu, temos o *Stellarium*, que é de uso livre, fácil de usar, com um desenho acessível até mesmo para crianças pequenas - http://www.stellarium.org/

**6.4 As fases da Lua em um espaço tridimensional: considerações para seu ensino**

Familiarizados com a interpretação e descrição das fases da Lua a partir da visão situada sobre a superfície terrestre, possivelmente, encontremos situações que não possam ser explicadas com precisão, ou que as ideias que foram definidas não sejam suficientes para explicar algumas das observações ou registos feitos. É possível que, por meio de observações, tenha-se compreendido que, quando a Lua está na sua fase de cheia, o Sol, provavelment,e será encontrado na extremidade oposta do céu. No entanto pode ser também que o aluno não possa explicar "por que a Terra não a encobre, se sabemos que é muito maior do que a Lua..."

Situações como essa nos fazem pensar ou construir um modelo que explique as observações. Ou ainda, é necessário entender como determinado modelo apresentado pelo professor, por exemplo, o atual modelo científico, explica muito bem aquelas questões que aparecem a partir das observações conhecidas. A inclusão deste novo modelo requer a mudança do lugar do observador, que deverá se afastar da superfície da Terra e observar o sistema STL de um ponto de vista no espaço distante. Com base em nossas pesquisas dos últimos anos, acreditamos que não seja tão fácil reconhecer esta mudança de ponto de vista, por isso, é preciso tornar explícita a mudança de perspectiva.

Para as diversas indagações já realizadas, reconhecemos que é provável que, no início, voltem a aparecer explicações alternativas, posto que estas idéias são fortes, estão arraigadas e são funcionais.

Dadas essas duas questões apresentadas, é conveniente modelizar os fenômenos associados ao sistema STL para nos aproximarmos do modelo explicativo, utilizando elementos concretos, que representem os três corpos. Nós buscamos, assim, respeitar, para o caso de a Lua e da Terra, a relação entre suas dimensões e considerar, adequadamente, seus movimentos e velocidades. É oportuno, portanto, dedicar algum tempo para se familiarizar com esta perspectiva, convidando os alunos a mudar sua posição de observador, por exemplo, fazendo-os transladar ao redor do sistema para percebê-lo a partir de diferentes posições no espaço. O trabalho inicial somente com o sistema Sol-Terra poderia ser usado para explicitar algumas ideias que serão tomadas como válidas: por exemplo, que o Sol não se move ou que a Terra tem dois movimentos, conhecidos por rotação e translação sobre seu eixo. Colocar essas ideias "sobre a mesa" é útil para o momento de apresentar novas situações que façam os alunos entrar em conflito e, por exemplo, que tentem "salvar" a situação, deslocando a linearidade do Sol ou da Terra.

Para facilitar a passagem do lugar do observador, pode ser valioso "materializar" um observador fictício com um palito, por exemplo, e colocá-lo de pé sobre um determinado local da superfície de um globo terrestre. Com este recurso, pode-se tentar reconhecer que fase lunar seria vista pelo "palito-observador terrestre", ajudando os alunos a olhar de modo adequado.

Nesse cenário de trabalho, será válido que o professor faça intervenções para que as situações se tornem complexas e para que o uso do novo modelo explicativo possa responder às situações que possam surgir. Serão novas circunstâncias nas quais se pode propor aos alunos que interpretem os registros e análises do fenômeno realizadas a partir da perspectiva localizada sobre a Terra, a fim de validar, ou não, com base nesse modelo, suas interpretações iniciais. As observações que foram feitas sobre os movimentos do Sol e da Lua, ligadas à discussão acima sobre o "arcos" aparentes que os astros descrevem no céu, valem como informação para pensar sobre a localização e movimentos dos objetos celestes em nosso modelo externo.

O trabalho com um modelo concreto que simula a explicação permite-nos introduzir o fenômeno dos eclipses que ocorrem no sistema STL, e possibilita-nos comparar esses eventos astronômicos com algumas das fases da Lua e discutir as frequências de cada um dos fenômenos.

**6.5 Representação bidimensional e explicação do fenômenos tridimensional**

As explicações dos fenômenos discutidos que podem ser alcançadas com a inclusão do modelo científico obriga-nos a pensar sobre os movimentos desses corpos em mais de um plano, ou seja, que os astros descrevem movimentos em planos diferentes, mas com pontos em comum. Esta condição requer não nos limitarmos aos movimentos bidimensionais e suas correspondentes representações planas.

No entanto, apesar de viver em um mundo tridimensional, as experiências que os alunos, muitas vezes, têm, em situações escolares, são, essencialmente, bidimensionais.

Em parte é por isso que se torna muito complexo para eles poderem representar no plano a explicação das fases como representações tridimensionais.

Ao mesmo tempo, como antecipamos no início do capítulo, há uma problemática na tridimensionalidade e em sua *representação* no plano, seja este o plano da lousa ou a folha do caderno, e é, particularmente, agravado naqueles esquemas que comumente são encontrados em livros didáticos. As imagens usadas para ilustrar as fases da Lua, pelo fato de serem, na maioria dos cados, bidimensionais, não são, suficientemente explicativas. Isso dificulta muito a interpretação de um fenômeno tridimensional.

Este último aspecto aumenta a complexidade do estudo das fases da Lua e a compreensão, por parte dos alunos, do modelo explicativo. Tais fatores tornam necessária a criação de oportunidades para analisar e discutir a validade dos diferentes tipos de representação.

**Sobre os autores:**

**Alejandro Gangui** é formado em Ciências Físicas pela Universidade de Buenos Aires, PhD em Astrofísica pela *International School for Advanced Studies*, Trieste, Itália. Atualmente, é investigador do Conicet, no Instituto de Astronomia e Física do Espaço, e professor na Universidade de Buenos Aires, Argentina.
Contato: *gangui@df.uba.ar*

**Esteban Dicovskiy** é formado em Ciências Físicas (Professor) pelo Instituto Superior da Universidade Tecnológica. Atualmente, é professor em Institutos de Formação Docente da cidade de Buenos Aires.
Contato: *esteban_dico@yahoo.com.ar*

**Maria C. Iglesias** é formada em Ciências Biológicas (Professora) pela Universidade de Buenos Aires, Diplomada em Construtivismo e Educação, pela Faculdade Latinoamericana de Ciências Sociais. Atualmente, é professora na Universidade de Buenos Aires, assessora em Ciências Naturais para a escola primária argentina, e autora de livros didáticos sobre Ciências Naturais.
Contato: *maryiglesias@gmail.com*

**Referências**

BAXTER, J. Children's understanding of familiar astronomical events. *International Journal of Science Education,* v.11, 1989, p. 502-513.

CAMINO, N. Aprender a imaginar para comenzar a comprender. Los «modelos concretos» como herramientas para el aprendizaje en astronomia. *Alambique - Didáctica de las Ciencias Experimentales,* v.42, 2004, p. 81-89.

DICOVSKIY, E.; IGLESIAS, M.; KARASEUR, F.; GANGUI, A.; CABRERA, J.; GODOY, E. El problema de la posición del observador y el movimiento tridimensional en la explicación de las fases de la luna en docentes de primaria en formación. *Actas de III Jornadas de Enseñanza e Investigación Educativa en el Campo de las Ciencias Exactas y Naturales*, Edición en CD-ROM, A. Vilches et al. (comp). Facultad de Humanidades y Ciencias de la Educación, Universidad Nacional de La Plata, 2012.

GANGUI, A.; IGLESIAS, M.; QUINTEROS, C.. Indagación llevada a cabo con docentes de primaria en formación sobre temas básicos de Astronomía. *Revista Electrónica de Enseñanza de las Ciencias*, v.9, n.2, 2010, pp. 467-486. Disponível em: http://saum.uvigo.es/reec/volumenes/volumen9/ART10_Vol9_N2.pdf . Acesso em mar./2013

IACHEL, G.; LANGHI, R.; SCALVI, R.M.F. Concepções alternativas de alunos do ensino médio sobre o fenômeno de formação das fases da lua. *Revista Latino-Americana de Educação em Astronomia* - RELEA, n. 5, 2008, p. 25-37.

IGLESIAS, M.; DICOVSKIY, E.; KARASEUR, F.; CABRERA, J.; GODOY, E.; GANGUI, A. La explicación de las fases de la luna en docentes de primaria en formación: aportes para la reflexión, en *Simposio de Investigación en Educación en Física*. 2012. Universidad Nacional de la Patagonia "San Juan Bosco".

KARASEUR, F. Los obstáculos en la comprensión de los fenómenos astronómicos en el uso de libros de texto. In: *III Jornadas Nacionales y I Latinoamericanas de Investigadores/as en Formación en Educación.* Facultad de Filosofía y Letras. Universidad de Buenos Aires, 2012.

LANCIANO, N. Ver y hablar como Tolomeo y pensar como Copérnico. *Enseñanza de las Ciencias*, v.7, n.2, 1989. p. 173-182.

MARTÍNEZ SEBASTIÀ, B. La enseñanza/aprendizaje del modelo Sol-Tierra: Análisis de la situación actual y propuesta de mejora para la formación de los futuros profesores de primaria. *Revista Latino-Americana Educação em Astronomia – RELEA*, n. 1, 2004. p. 7-32.

SAMARAPUNGAVAN, A.; VOSNIADOU, S.; BREWER, W.F. Mental models of the Earth, Sun and Moon: indian children's cosmologies. *Cognitive Development,* v.11, 1996. p. 491-521.

SCHOON, K. J. Student´s alternative conceptions of Earth and space. *Journal of Geological Education,* v.40, 1992. P. 209-214.